\documentclass[twocolumn,epsfig,pre]{revtex4}

\usepackage{graphics}
\usepackage{graphicx}
\usepackage{epstopdf}
\usepackage{amssymb}
\usepackage{amsmath}
\usepackage{hyperref}
\usepackage{latexsym}
\usepackage{color}
\usepackage{soul}

\begin{document}
	
	\renewcommand{\thefootnote}{\fnsymbol{footnote}}
	\renewcommand{\theequation}{\arabic{section}.\arabic{equation}}
	
	%\title{Effect of energy distribution shape on the heteropolymer collapse transition}
	\title{Universality and identity ordering in heteropolymer coil-globule transition}

	\author{Thoudam Vilip Singh}
	%\email{thoudamsinghsingh@gmail.com}	
	\author{Lenin S.~Shagolsem}
	\email{slenin2001@gmail.com}
	\affiliation{Department of Physics, National Institute of Technology Manipur, Imphal, India} 
	
	\date{\today}

\begin{abstract}

\noindent The coil-globule transition of an energy polydisperse chain, a model heteropolymer system where the number of monomer species is as large as the total number of monomers, is studied by means of computer simulations. In this study, we systematically explore the consequences of having different functional form and variance of the energy distribution on the coil-globule transition in general. 
In particular, considering Gaussian (G) and uniform (U) distributions, the effect of varying polydispersity index, $\delta$, on the transition temperature $\theta^\ast$, chain size, internal structure and spatial organization of monomers in the globule, and kinetics of the folding are addressed. 
It is found that the transition temperature of the model heteropolymer is lower than that of the homopolymer counterpart, and $\theta^\ast$ increases with $\delta$ (both G and U) and $\theta^\ast({\rm U}) < \theta^\ast({\rm G})$ consistently. The results of our study suggest that $\theta^\ast$ is governed by the most probable value (rather than the width) of the pair-wise energy distribution. Interestingly, the nature of the collapse transition turns out to be universal, i.e., when scaled properly (irrespective of the functional form and variance) all the swelling curves fall on a master curve and it is well described by the same scaling form of the homopolymer counterpart. 
However, following quenching, the transition from coil to globule is relatively fast for heteropolymer (with no significant difference between G and U systems, and no significant $\delta$ dependence within the considered range). 
On the other hand, internal organization in the collapsed state, quantified through mean contact probability, show distinct scaling regimes. Also we observe segregation of monomers based on their identities which is more pronounced in the case of uniform distribution.  

\end{abstract}

\maketitle

%%\tableofcontents

%%%%%%%%%%%%%%%%%%%%%%%%%%%%%%%%%%%%%%%%%%%%%%%%%%%%%%%%%%%%%%%%%%%%%
\section{Introduction} \label{sec: intro}

A polymer chain in an extended or coil conformation can undergo structural transition to a compact globular structure and vice versa upon changing its environment (like temperature and PH) which is often seen in a number of biomacromolecules, e.g.~protein folding and denaturation process, and in synthetic polymers.\cite{Sherman,Frerix,Tanaka0,Kita,Xuuu,Nakata} 
The coil-globlue transition was first predicted by Stockmayer in 1960 and observed experimentally by Tanaka et al.~in 1979.\cite{Stockmayer,Tanaka1,Tanaka2} 
It is clear by now that the observed structural transition is a result of the contest between the contribution of intramolecular interactions of the monomers and the loss of entropy due to swelling.\cite{Lifshitz,Grosberg1,Grosberg2,Grosberg3,Grosberg4,Grosberg5} 
For fully flexible polymer chains, such transition is of second order and for semi-flexible or stiff polymers it is of first order.\cite{Grosberg6,Post,Rampf,Noguchi,Bastolla}
In addition to its fundamental interest, the phenomena of coil-globule transition is also relevant to protein folding and DNA condensation,\cite{Go,Dill,Onuchic,Bloomfield} network collapse, etc., and thus it has been a subject of intense research.\cite{Lesage,MaedaY,DingY,Birshtein,Tanaka,Wittkop,Polson,Maffi,Holehouse} \\

Biopolymers have multi-species domains or residues, e.g.~proteins and DNAs/RNAs have different residues which in a coarse-grained picture represent energetically different regions. Thus proteins are not simple homopolymer, but rather a finite-size heteropolymers. Since real protein sequences statistically look like random sequences the theoretical works that provide many aspects of the folding process employ random heteropolymer (RHP) model,\cite{Bryngelson,Shakhnovich,shakhnovich2006,Socci,Pande} where the simplest model consider two kinds of residues (i.e.~hydropbic and hydrophilic amino acids) randomly distributed along the chain and in more general case it goes up to 20 amino acids. 
Describing the physics of heteropolymer freezing (as a simple approximation to protein folding) relies on random energy model (REM), developed by Derrida \cite{Derrida} to study spin glass system.\cite{Pande1997} 
The qualitative study of forming a unique structure in polypeptide chains and the enumeration of compact states of copolymers were carried out validating the use of REM in heteropolymers, and even highlighting its potential in understanding the evolution of proteins.\cite{Shakhnovich,Shakhnovich1}
Although RHP model gives good insight into the fundamentals of heteropolymer freezing, it lacks some important features, e.g.~cooperative folding leading to native structure of the protein, and this leads to the idea of designed heteropolymer with the general requirements for sequences to be protein-like.\cite{shakhnovich2006} \\

The RHP also represent intrinsically disordered proteins (IDP) which do not have a well defined native structure. Despite lacking a unique state, IDP plays a vital role in biological condensates of cells, ranging from cell signalling, transduction, cell physiology to diseases.\cite{McCarty, Dyson1, Shin, Banani} An interesting phenomenon observed in these disordered proteins is coacervation, a phase condensation leading to a liquid-liquid phase separation.\cite{Patel, Molliex} 
The known principles of polymer physics have been applied to study liquid-liquid phase separation of proteins and nucleic acids because they share familiar characteristics of the well-accepeted phase behaviour of polymer solutions.\cite{Quiroz, Brangwynne, Lin, Dignon} \\

In studying such complex systems, computer simulations, in addition to experiments and theory, have been of immense help in revealing many general aspects of the coil-globule transition.\cite{Dai,Polson,Steinhauser,Wittkop,Oever} 
There already exists a large volume of literature in the homo/hetero-polymer coil-globule transition, and yet, in this study, we reconsider the collapse transition of an energy polydisperse chain, a generic model heteropolymer, to highlight the role of the underlying energy distribution and its consequence on the coil-globule transition, in general, by means of molecular dynamics simulations.   
To this end, we consider two types of energy distributions, namely, Gaussian and uniform distributions, and investigate the role of distribution shape on the coil-globule transition temperature, and the effect of variance is also studied. (Use of other type of distribution function, e.g., log-normal distribution is reported in reference~\cite{vilip2021}.) In addition to this, we also address whether the nature of collapse transition depends on the distribution function or not. Another pertinent question is how rapid the globule formation is compared to the reference homopolymer system (of same molecular weight and same mean interaction energy). 
One of the interesting aspect is the internal organization and identity ordering of monomers in the globular state which we study by means of contact map analysis. A contact map provides information of the preferred monomer-monomer interactions and it has been used to study the 3D structural organization of proteins in folded states.\cite{Vendruscolo, Shoemaker, Dyson} 
Both the folding of proteins and the complex three dimensional organizations of chromosomes have biomolecular machinery which has yet to be properly understood like the gene activity and the functional state of a living cell.\cite{Chiariello} \\

The remainder of the paper is organized as follows. Model and simulation details are presented in section~\ref{sec: model-description}. While in section~\ref{sec: results and discussion}, the results on the coil-globule transition and $\theta$-temperature, internal organization and energetics, and folding kinetics are discussed. Finally, we conclude the paper in section~\ref{sec: summary}. 

%\clearpage

%%%%%%%%%%%%%%%%%%%%%%%%%%%%%%%%%%%%%%%%%%%%%%%%%%%%%%%
%%%%%%%%%%%%%%%%%%%%%%%%%%%%%%%%%%%%%%%%%%%%%%%%%%%%%%%

\section{Model and simulation details}
\label{sec: model-description}

The polymer chains are simulated using coarse-grain bead-spring model of Kremer and Grest,\cite{kremer_JCP_92} where a polymer chain is represented by beads connected with springs. All the pairwise interactions in the system is modeled via Lennard-Jones (LJ) potential,
\begin{equation}
U_{\tiny{_{\rm LJ}}}(r) = 4\varepsilon_{ij}\left[(\sigma/r)^{12}-(\sigma/r)^{6}\right]~,
\label{eqn: LJ-potential}
\end{equation}
where $r$ is the separation between a pair of particles, $\sigma$ the monomer diameter, and $\varepsilon_{ij}=\sqrt{\varepsilon_i\varepsilon_j}$ the interaction strength between a pair of particles $i-j$ (following Lorentz-Berthelot mixing rule~\cite {Berthelot}). The LJ potential is cut-off and shifted to zero at $r=2.5\sigma$. The monomer connectivity along the chain is assured via finitely extensible, nonlinear, elastic (FENE) springs~\cite{grest_kramer_PRA_33} represented by the potential, 
\begin{equation}
U_{\tiny{_\text{FENE}}} = \left\{
\begin{array}{l l}
-\frac{kr_0^2}{2}\ln\left[1-\left(r/r_0\right)^{2}\right]~, & \quad r<r_0 \\ %\nonumber\\ 
\infty~, & \quad r\ge r_0
\end{array} \right.
\label{eqn: fene-potential}
\end{equation}
where $r$ is the separation of neighboring monomers along the chain. The spring constant $k=30\varepsilon/\sigma^2$ and the maximum extension between two consecutive monomers 
along the chain $r_0=1.5\sigma$. The above values of $k$ and $r_0$ ensure that the chains avoid bond crossing and very high frequency modes~\cite{kremer_JCP_92, Kremer}.  

\begin{figure}[ht]
	\begin{center}
		\includegraphics[width=0.45\textwidth]{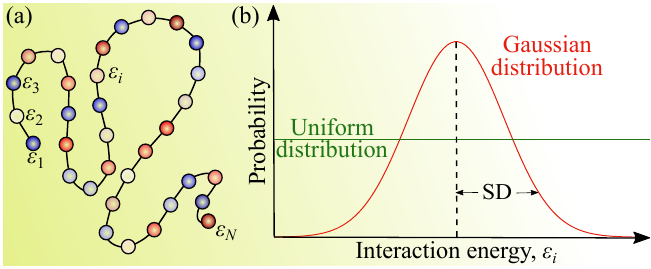}
		\caption{(a) A cartoon of energy polydisperse chain, where each monomer is identified by the interaction parameter $\varepsilon_{i}$ randomly drawn for a given distribution. (b) Shape of the Gaussian and uniform distributions.}
		\label{fig: schematic-1}
	\end{center}
\end{figure}

For this study we consider two types of chain differing in their interaction energy distribution, namely, Gaussian (G) and uniform (U) energy distributions described in the following. As sketched in figure~\ref{fig: schematic-1}, each monomer is assigned an interaction energy, $\varepsilon_i$, drawn randomly from a Gaussian distribution (refer afterwards as G-system) defined as 
\begin{equation}
P(\varepsilon_i)= 
\frac{1}{\sqrt{2 \pi \left( \text{SD} \right)^2}} \text{exp} \left[-\frac{\left(\varepsilon_i - \langle\varepsilon \rangle \right)^2}{4 \left( \text{SD} \right)^2} \right]~, 
\label{eqn: gaussian-dist}
\end{equation}
with SD the standard deviation and $\langle \varepsilon \rangle = (\sum_{i=1}^{N} \varepsilon_i)/N$ the mean; and uniform distribution defined as 
\begin{equation}
P(\varepsilon_i)= \begin{cases}
1/\left(n - m \right), \quad  ~~\text{for} \quad m \le{\varepsilon_i} \le{n}\\
0,  \quad \qquad \quad \qquad  \text{elsewhere.}
\end{cases}
\label{eqn: uniform-dist}
\end{equation}
For uniform distribution (refer afterwards as U-system) in the range $[m,n]$ the mean and the SD are, respectively,   
\begin{equation}
\langle\varepsilon \rangle = \frac{m + n}{2}~~{\rm and}~~
\text{SD} = \frac{n - m}{2\sqrt{3}}.
\label{eq: uniform-dis-SD}
\end{equation}

In this study, for both distribution types, the mean is fixed at $\langle\varepsilon\rangle = 2.5$  and the polydispersity \rm{is varied} by changing the SD. Also for the reference homopolymer chain, we set the value of pair-wise interaction at $\langle \varepsilon \rangle$. The degree of polydispersity is quantified through polydispersity index $\delta~(={\rm SD}/{\rm mean})$, and for uniform distribution it is  
\begin{equation}
\delta = \frac{n - m}{\sqrt{3}\left(m + n \right)}.
\end{equation}
%Typically $\delta$ is varied upto 13\%. 
The model considered here belongs to the independent interaction model (where the number of monomer spercies is as large as the total number of monomers) which is used in modeling heteropolymer or disordered proteins.\cite{Pande} 
All the physical quantities used here are expressed in LJ reduced units,\cite{Frenkel,Allen} where $\sigma$ and $\varepsilon$ are the basic length and energy scales respectively. The reduced temperature $T$ and time $t$ are defined as $T = k_{\rm_B}T_0/\varepsilon$ and $t = t_0/\tau_{_{\rm LJ}}$, where $\tau_{_{\rm LJ}}=\sigma\sqrt{m/\varepsilon}$ represents the LJ time unit, and $k_{\rm_B}$, $T_0$, $t_0$, and $m$ are the Boltzmann constant, absolute temperature, real time, and mass, respectively. \\ 

The molecular dynamics (MD) simulations were carried out using Langevin dynamics\cite{Allen,Frenkel} where the equations of motion are given by  
\begin{equation}
m_i \frac{{\rm d}^2 {\bf r}_i}{{\rm d}t^2} + \gamma \frac{{\rm d} {\bf r}_i}{{\rm d}t} = -\frac{\partial U}{\partial {\bf r}_i} + {\bf f}_i(t)~,
\label{eqn: langevin}
\end{equation}
where ${\bf r}_i$ and $m_i$ are the position and mass of particle $i$, respectively, and $\gamma$ being the friction coefficient which is taken to be the same for all particles, $U=U_{\tiny{_{\rm LJ}}}+U_{\tiny{_\text{FENE}}}$ is the potential acting on monomer $i$ of the polymer chain, and ${\bf f}_i$ are random external forces which follows the relations: $\left\langle {{\bf f}_i}(t)\right\rangle=0$ and $\left\langle {{\bf f}_i}^{k}(t){{\bf f}_j}^{l}(t')\right\rangle=2\gamma m_i k_{\rm_B}T\delta_{ij}\delta_{kl}\delta(t-t')$ where $k$ and $l$ denotes the Cartesian components. 
The friction coefficient $\gamma=1/\tau_d$, with $\tau_d$ the characteristic viscous damping time which we fixed at $50$ and it determines the transition from inertial to overdamped motion. The chosen value of $\gamma$ gives the correct thermalization for our study. The equations of motion are integrated using velocity-Verlet scheme with a time step of $\delta t_0 = 0.005 \tau_{_{\rm LJ}}$. \\

In this study, a single polymer chain is tethered at the center of a cubic simulation box which is periodic in all directions. Various chain lengths in the range $64\le N \le 2048$ are considered and depending on the chain size simulation box size varies in the range $100\sigma \le L \le 250\sigma$. 
Taking a chain relaxed at high temperature (where it assumes extended conformation) we quenced the chain to various lower temperatures and equilibrate at the temperature for sufficient amount of time where the thermodynamic quantities, e.g., total energy remains constant with time and then followed by production runs where various quantities of interest are measured. Typical relaxation time varies in the range $1\times 10^6 - 1\times 10^7$ MD steps, and production runs last between $5 \times 10^6 - 1 \times 10^8$ MD steps depending on the chain length. The results reported here are obtained by averaging over 20 replicas with each replica having a different set of interaction parameter $\varepsilon_i$. All the simulations were carried out using LAMMPS code.\cite{Plimpton}

%%%%%%%%%%%%%%%%%%%%%%%%%%%%%%%%%%%%%%%%%%%%%%%%%%
%%%%%%%%%%%%%%%%%%%%%%%%%%%%%%%%%%%%%%%%%%%%%%%%%%
%\section{Results}
%\label{sec: results}
%%%%%%%%%%%%%%%%%%%%%%%%%%%%%%%%%%%%%%%%%%%%%%%%%%
%%%%%%%%%%%%%%%%%%%%%%%%%%%%%%%%%%%%%%%%%%%%%%%%%%
\section{Results and discussion} \label{sec: results and discussion}

\begin{figure}[ht]
	\begin{center}
		\includegraphics[width=0.47\textwidth]{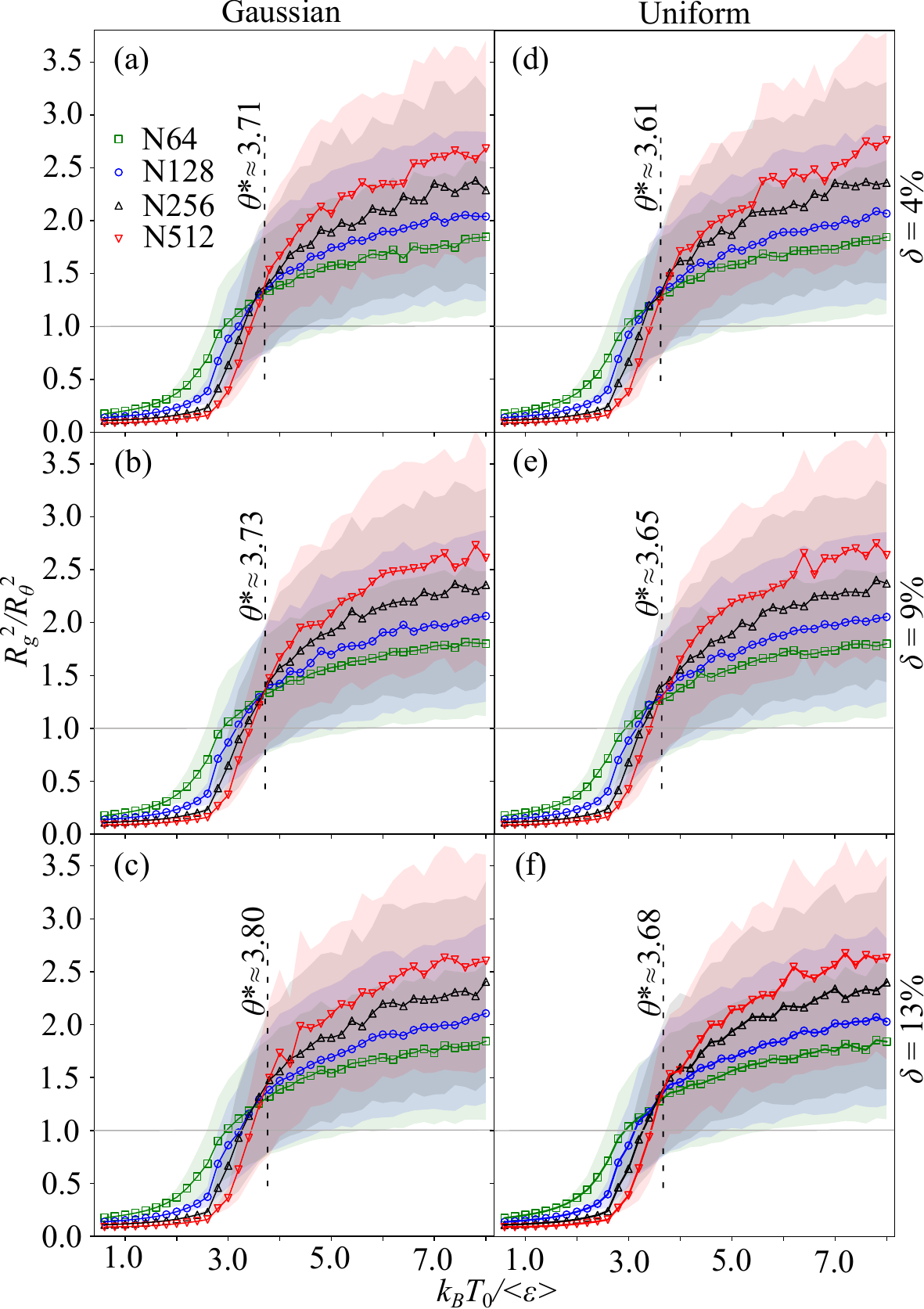}
		\caption{Plot of swelling ratio, $\alpha^2=R_g^2/R_{\theta}^2$, as a function of temperature (rescaled by the mean energy $\langle \varepsilon \rangle$) for energy polydisperse chains (Left/Right panel: Gaussian/Uniform distribution) of different chain length $N$ shown for three different values of $\delta$ indicated in the figure. The shaded regions indicate the rms error obtained from different replicas.}
		\label{fig: scaling-Y}
	\end{center}
\end{figure}

\subsection{Collapse Transition}
\label{sec: theta-temp}

\begin{figure}[]
	\begin{center}
		\includegraphics[width=0.47\textwidth]{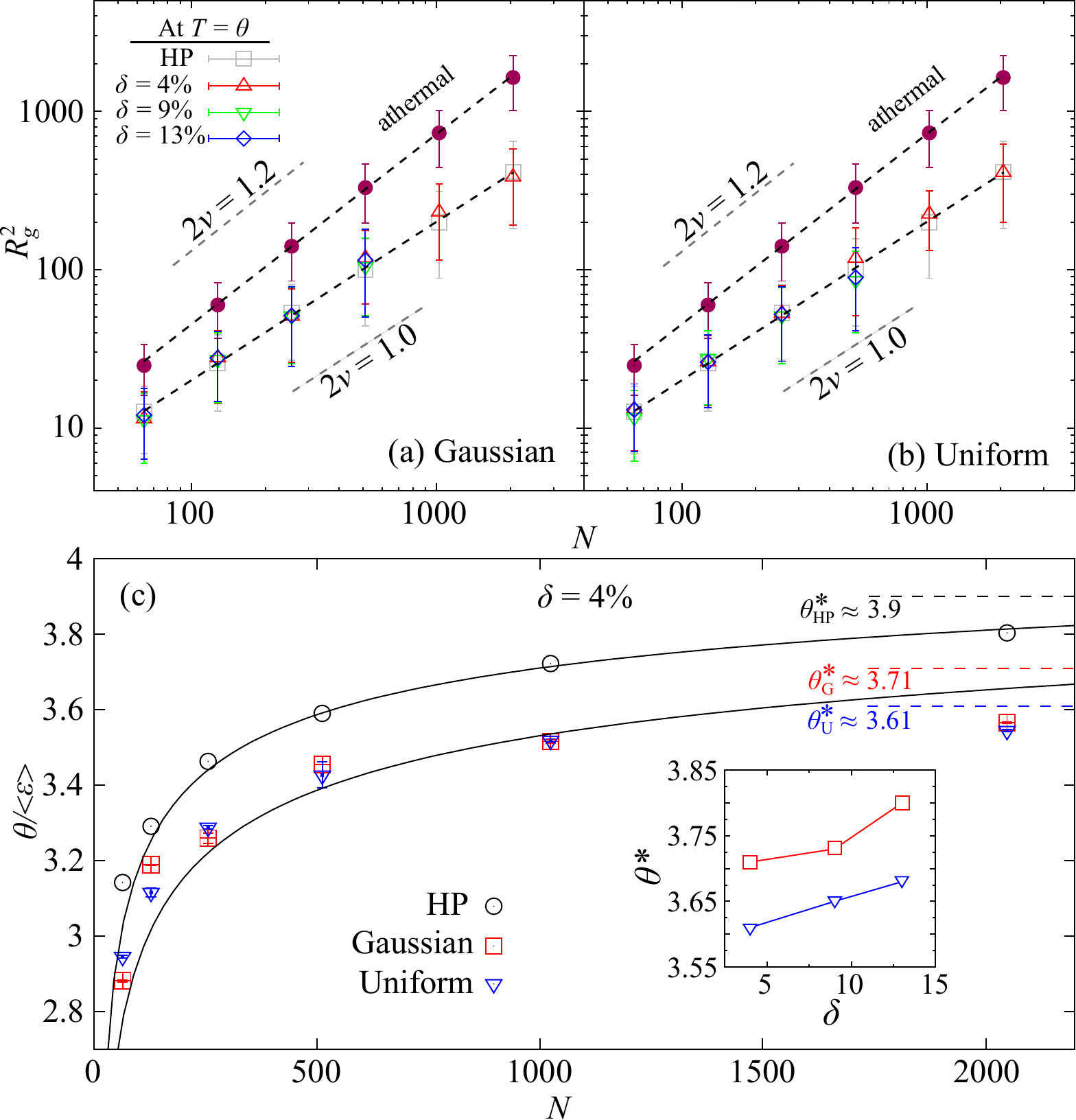}
		\caption{$R_g^2$ vs $N$ at $\theta$-temperature for both homopolymer (HP) and (a) Gaussian and (b) uniform energy polydisperse chains at polydispersity index, $\delta = 4\%,~9\%,~13\%$. For comparison, the chain size under athermal condition is also obtained. Within errorbar the Flory exponent $\nu\approx 0.5$ ($\theta$ condition) and $\nu \approx 0.6$ (athermal consition). (c) Plot of $\theta/\langle\varepsilon\rangle$ as a function of $N$ for energy polydisperse chains at $\delta=4\%$. Solid lines represent the fitted curves using $\theta \approx \theta^\ast (1-1/\sqrt{N})$. Inset of figure (c) is the plot of $\theta^\ast$ (see figure~\ref{fig: scaling-Y}) as a function of $\delta$.}
		\label{fig: N-theta}
	\end{center}
\end{figure}
\begin{figure}[!h]
	\begin{center}
		\includegraphics[width=0.47\textwidth]{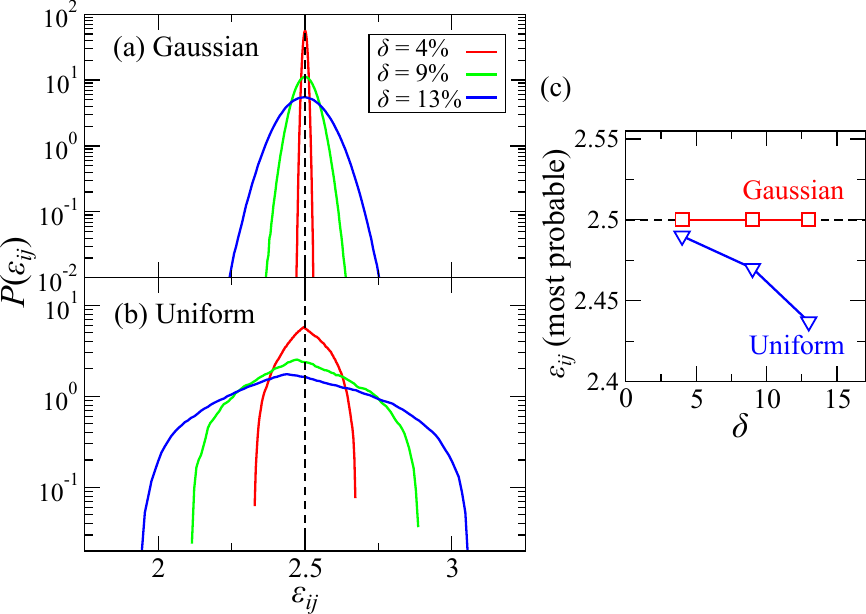}
		\caption{Distribution of $\varepsilon_{\rm ij}=\sqrt{\varepsilon_{\rm i}\varepsilon_{\rm j}}$ for (a) Gaussian and (b) Uniform distributions of $\varepsilon_{\rm i}$, see eqn.~\ref{eqn: gaussian-dist} and eqn.~\ref{eqn: uniform-dist}, at polydispersity index, $\delta~=~4\%,~9\%,~13\%$. Notice that for uniform distribution, the mean and most probable values do not coincide. (c) The most probable value, $\varepsilon_{\rm ij}^{\rm mp}$, of the distributions shown as a function of $\delta$ (in \%), and the mean value is indicated by the horizontal dashed line at 2.5.}
		\label{fig: eps-ij-distribution}
	\end{center}
\end{figure}

Upon changing temperature (or solvent quality), a polymer chain can transform its state from an extended coil conformation to a collapsed globule and vice-versa. At a particular temperature called $\theta$-temperature, the second virial coefficient vanishes and it leads to the conformation of a real chain adopting that of an ideal chain and thus the chain size, $R_{\theta} \sim N^{1/2}$, with $N$ being the number of monomers in the chain.\cite{Rubinstein} 
The collapse transition is driven by the balance of net monomer-monomer interaction and the conformational entropy of the chain.\cite{Grosberg6} For instance, at a low temperature the effective monomer-monomer interaction is more attractive relative to monomer-solvent interaction and so the chains adopt a compact globular shape. However with decreasing chain size, the conformational entropy also decreases. The free energy of the polymer chain, in terms of swelling ratio, $\alpha$, can be written as    
\begin{equation}
F(\alpha)/k_{\rm B}T = F_{\rm el}(\alpha) + F_{\rm int}(\alpha)~,
\label{eqn: free-energy} 
\end{equation}
with $\alpha^2 = R_g^2/R_{\theta}^2$ where $R_g^2$ is the mean-square radius of gyration, and the elastic free energy and the interaction free energy (in mean-field approximation) are 
\begin{eqnarray}
F_{\rm el}(\alpha) &\approx& A \left[ \alpha^{-2} + 2\ln \alpha \right],~{\rm and}\\
F_{\rm int}(\alpha) &\approx& N \left( Bn + C n^2 \right)
\label{eqn: free-energy-1} 
\end{eqnarray}
respectively, with $n=N/V$ as the monomer density, $A$ being a constant, $B$ and $C$ are the second and third virial coefficients respectively. Minimization of equation~\ref{eqn: free-energy} w.r.t.~$\alpha$ gives the Flory formula for the swelling coefficient of the chain: 
\begin{equation}
\alpha^5-\alpha^3 = {\rm const.}~ N^{1/2}B(T)
\label{eqn: flory-formula}
\end{equation}
which captures the main feature of coil-globule transition and serves as a basis for understanding the phenomena.\cite{Lifshitz,Grosberg1,Grosberg2,Grosberg3,Grosberg4,Grosberg5,Birshtein} A more detailed treatment of the same are due to Sanchez.\cite{Sanchez}. 
In the above equation, the second virial coefficient, $B(T)$, which gives the excluded volume, $v$, contribution is related to the pair-wise interaction potential, $u_{ij}~(=U_{\rm LJ})$, as
\begin{eqnarray}
v &=& B(T) = \int \left( 1 - e^{u/{k_{\rm B}T}} \right) {\rm d}^3r~ \nonumber \\
{\rm or}~~ v &\approx& \left( 1 - \frac{\theta^\ast}{T} \right)\sigma^3~,
\end{eqnarray}
in which the effective transition temperatute, $\theta^\ast$, is given by 
\begin{equation}
\theta^\ast \approx -\frac{1}{k_{\rm B}\sigma^3} \int_{\sigma}^{\infty} u_{ij} r^2 {\rm d}r~.
\label{eqn: theta-temp}
\end{equation}
When $T=\theta^\ast$, the exclded volume $v=0$ and hence we recover the value of Flory exponent, $\nu=3/5$. From equation~\ref{eqn: theta-temp}, it is clear that the $\theta$-temperature depends explicitly on the form of pair potential which in turn depends on the distribution of pair interaction parameter, $\varepsilon_{ij}$, see equation~\ref{eqn: LJ-potential}. \\

In figure~\ref{fig: scaling-Y}, we plot the swelling ratio, $\alpha$, as a function of rescaled temperature for chains of various sizes shown at three different values of polydispersity index, $\delta$. The temperature at which $\alpha\approx 1$ (denoted by $\theta$) differs for different chain length, see figure~\ref{fig: N-theta}, for the chain size scaling and variation of $\theta$ as a function of $N$. The $\theta$-temperature varies with $N$ as $\theta \approx \theta^\ast(1-1/\sqrt{N})$, where $\theta^\ast$ is the ideal or true transition temperature,\cite{Rubinstein} which one can achieve in the limit $N\gg 1$. At the transition temperature, the polymer chain behaves as an ideal chain and the plot of swelling ratio for different $N$ show a common intersection point at $T=\theta^\ast$ where the curvature changes.\cite{Steinhauser} The same analysis is carried out for the reference homopolymer system to determine $\theta(N)$ and $\theta^\ast$ (see SI). For the reference homopolymer system, we obtain $\theta^\ast \approx 3.9$ (see SI), while it is relatively low for energy polydisperse chains. 

The plot of $\theta^\ast$ vs polydispersity index $\delta$ is shown as inset in figure~\ref{fig: N-theta}(c). For both Gaussian and uniform systems the value of $\theta^\ast$ increases with increasing $\delta$. However, the value is consistently lower for the system with uniform energy distribution. Thus, the observed trend is (a) $\theta^\ast({\rm U}) < \theta^\ast({\rm G}) < \theta^\ast({\rm HP})$ at a given value of $\delta$, and (b) $\theta^\ast$ increases with increasing $\delta$ for both uniform and Gaussian energy polydisperse polymers. The observed increase of $\theta^\ast$ with $\delta$ is consistent with the result of earlier theoretical calculations where it is shown that the transition from a coil to random-globule is directly proportional to variance of the distribution.\cite{Pande1997} \\ 

Since $\theta^\ast$ is related to the interaction potential $u_{ij}$ of the system, see equation~\ref{eqn: theta-temp}, and $u_{ij} \propto \varepsilon_{ij}$ through equation~\ref{eqn: LJ-potential}, meaning that the transition temperature depends on the type of energy distribution. As we can see in figure~\ref{fig: eps-ij-distribution}, the distribution $P(\varepsilon_{ij})$ of $\varepsilon_{ij}=\sqrt{\varepsilon_i \varepsilon_j}$, i.e.~the convolution of two random variables $\varepsilon_i$ and $\varepsilon_j$, is also Gaussian for the Gaussian distribution of $\varepsilon_i$ (equation~\ref{eqn: gaussian-dist}) where the mean $\langle \varepsilon_{ij} \rangle$ and most probable $\varepsilon_{ij}^{\rm mp}$ values coincide, while for the uniform distribution (equation~\ref{eqn: uniform-dist}) $\varepsilon_{ij}^{\rm mp} < \langle \varepsilon_{ij} \rangle$, and $\varepsilon_{ij}^{\rm mp}$ shifts to a smaller value with increasing variance of the distribution, see figure~\ref{fig: eps-ij-distribution}(c). And at the same value of polydispersity index $\delta$, the distribution $P(\varepsilon_{ij})$ is relatively broad for uniform distribution. These observations suggest that the value of $\theta^\ast$ is dictated by the value of $\varepsilon_{ij}^{\rm mp}$ rather than the width of the distribution in the sense that $\varepsilon_{ij}^{\rm mp}({\rm U}) < \varepsilon_{ij}^{\rm mp}({\rm G})$ (although uniform distribution is broader) and thus $\theta^\ast({\rm U}) < \theta^\ast({\rm G})$. Moreover, for a given distribution (i.e.~Gaussian or uniform), the increase in $\theta^\ast$ with increasing $\delta$ indicates that, here, the transition temperature is dictated by the population of larger $\varepsilon_{ij}$. \\
%%%%%%%%%%%%%%%%%%%%%%%%%%%%%%%%%%%%%%%%%%%%%%%%%%
%%%%%%%%%%%%%%%%%%%%%%%%%%%%%%%%%%%%%%%%%%%%%%%%%%
%\section{Nature of collapse transition} 
%\label{sec: size-scaling} 
\begin{figure}[ht]
	\begin{center}
		\includegraphics*[width=0.47\textwidth]{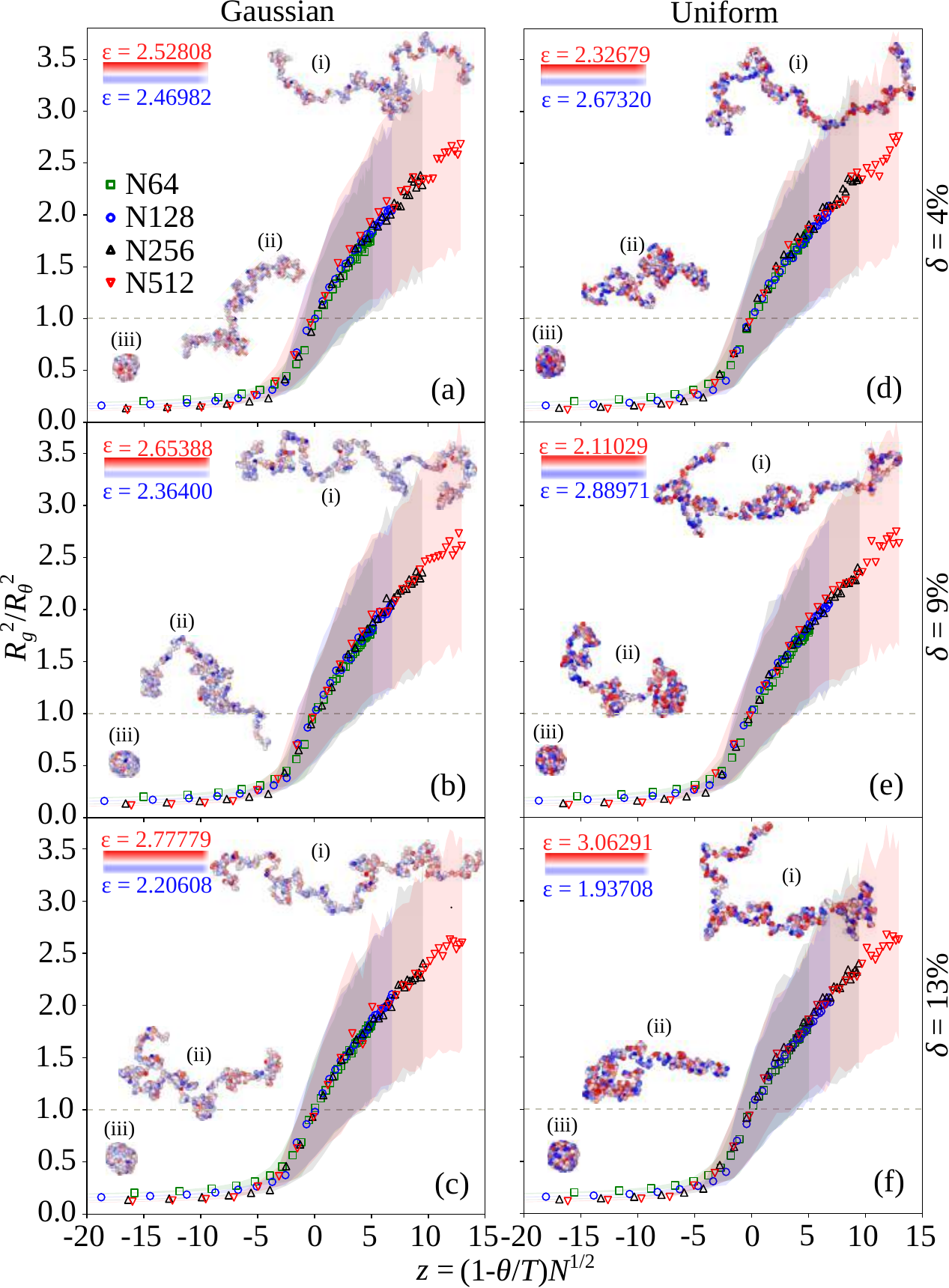} 
		\caption{Replot of swelling ration (see figure~\ref{fig: scaling-Y}) as a function of chain interaction parameter, $z=\left(1-\theta^\ast/T\right)N^{1/2}$, for Gaussian (Left-panel) and uniform (Right-panel) energy polydisperse chains. Polymer conformations shown are for (i) $z \approx 13$, (ii) $z \approx -20$, and (iii) $z \approx -169$. For homopolymer, see SI.}
	\label{fig: scaling-xy}
	\end{center}
\end{figure}

Finally, in order to see whether there is any qualitative difference on the nature of collapse transition between the two energy polydisperse systems, we replot the swelling curve shown in figure~\ref{fig: scaling-Y} as a function of interaction parameter $z=\left(1-\theta^\ast/T\right)N^{1/2}$. As we can see in figure~\ref{fig: scaling-xy}, irrespective of energy distribution types and polydispersity value, all the curves collapsed to a master curve and this is not different from the homopolymer collpase. Thus the collapse transition behavior is universal, and it shown that the chain swelling ratio $\alpha \approx |z|^{-1/3}$ for $T<\theta^\ast$ and $\alpha \approx z^{2\nu-1}$ for $T>\theta^\ast$ with $\nu$ being the Flory exponent.\cite{Rubinstein} In the following we discuss the results on the analysis of favorable contacts in the globular state.  

%\clearpage

%%%%%%%%%%%%%%%%%%%%%%%%%%%%%%%%%%%%%%%%%%%%%%%%%%
%%%%%%%%%%%%%%%%%%%%%%%%%%%%%%%%%%%%%%%%%%%%%%%%%%
\subsection{Internal structure and energetics} 
\label{sec: contact-analysis}

\begin{figure}[] 
	\begin{center}
		\includegraphics*[width=0.48\textwidth]{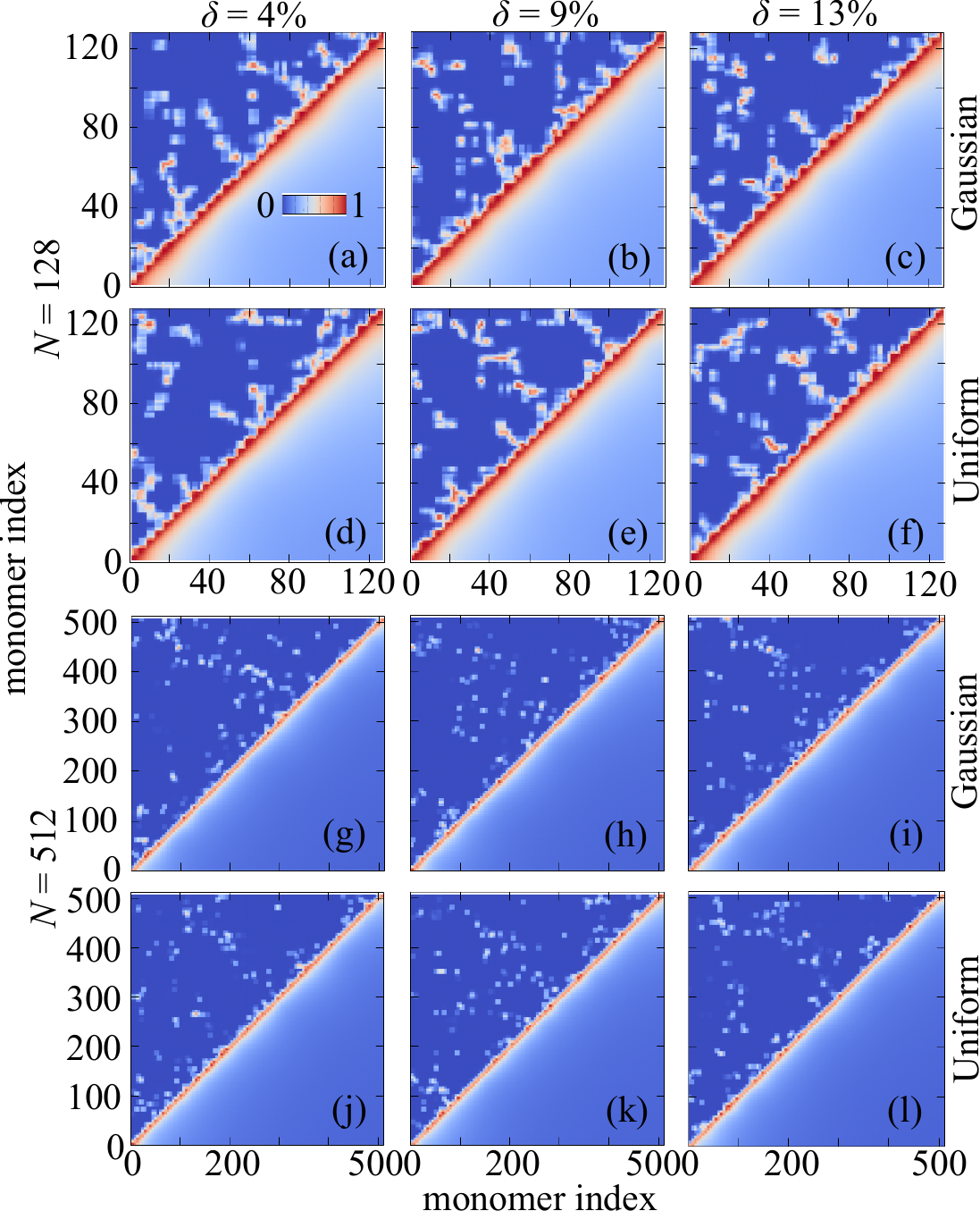}
		\caption{Contact maps of Gaussian, (a)-(c), and uniform, (d)-(f), energy polydisperse chains at $\delta=4\%,9\%,12\%$, respectively, for chain length $N$ = 128. In figure (g)-(l), similar contact maps for a larger chain, i.e., $N$ = 512 is shown. Contact maps are obtained from 5000 independent chain conformations at $T_0=1.0$ (or $z\approx -169$, see figure~\ref{fig: scaling-xy}), where it forms a well rounded globule.}
		\label{contact-map}
	\end{center}
\end{figure}

\begin{figure}
	\begin{center}
		\includegraphics*[width=0.47\textwidth]{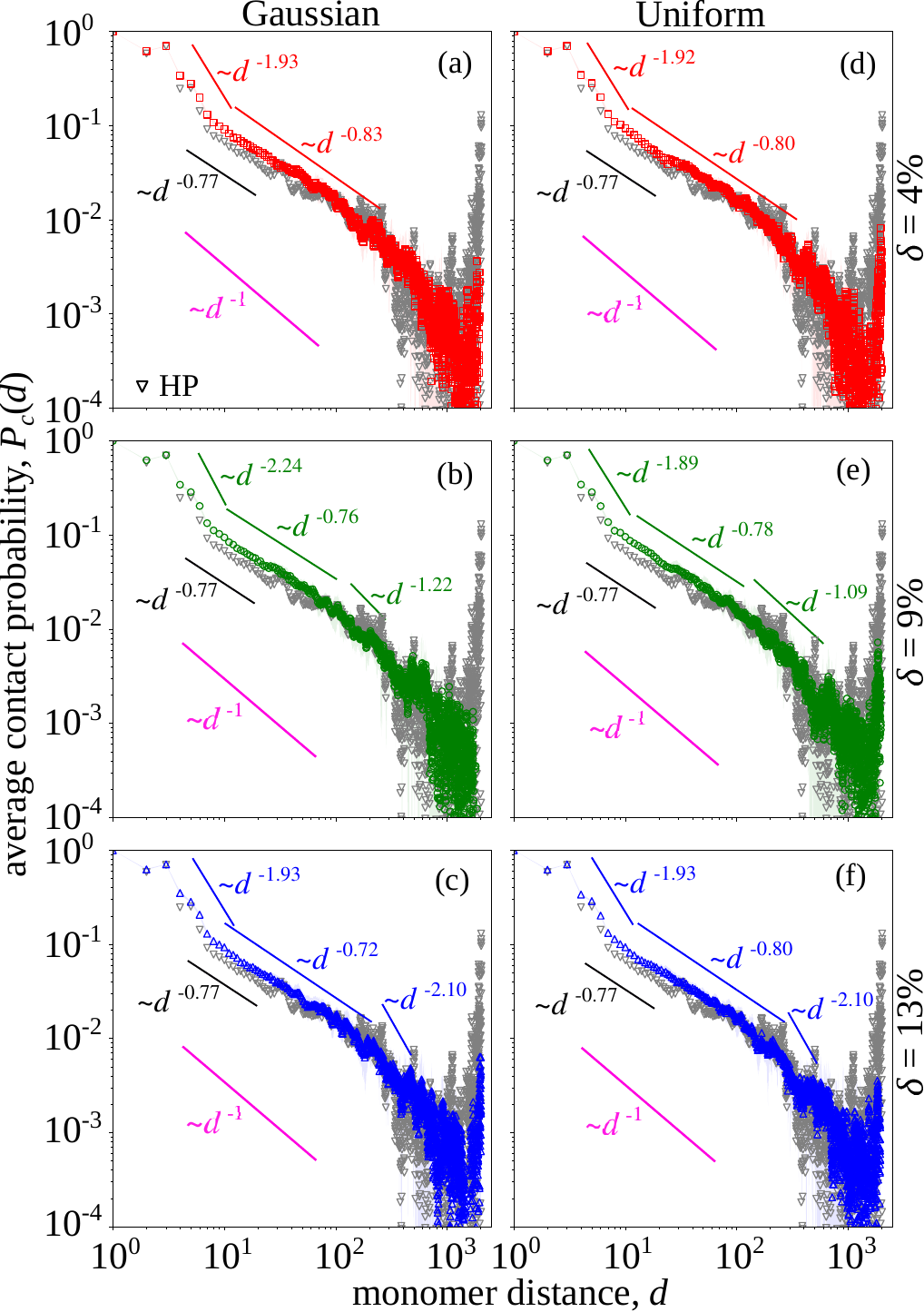}
		\caption{Contact probability of energy polydisperse chain ($N=2048$) shown for $\delta$ = 4\%, 9\%, 13\% (top to bottom) for Gaussian, figure~(a)-(c), and uniform, figure~(d)-(f), distributions respectively.}
		\label{fig: mean-contact-prob} 
	\end{center}
\end{figure}

\begin{figure*}[ht]
	\begin{center}
		\includegraphics*[width=0.85\textwidth]{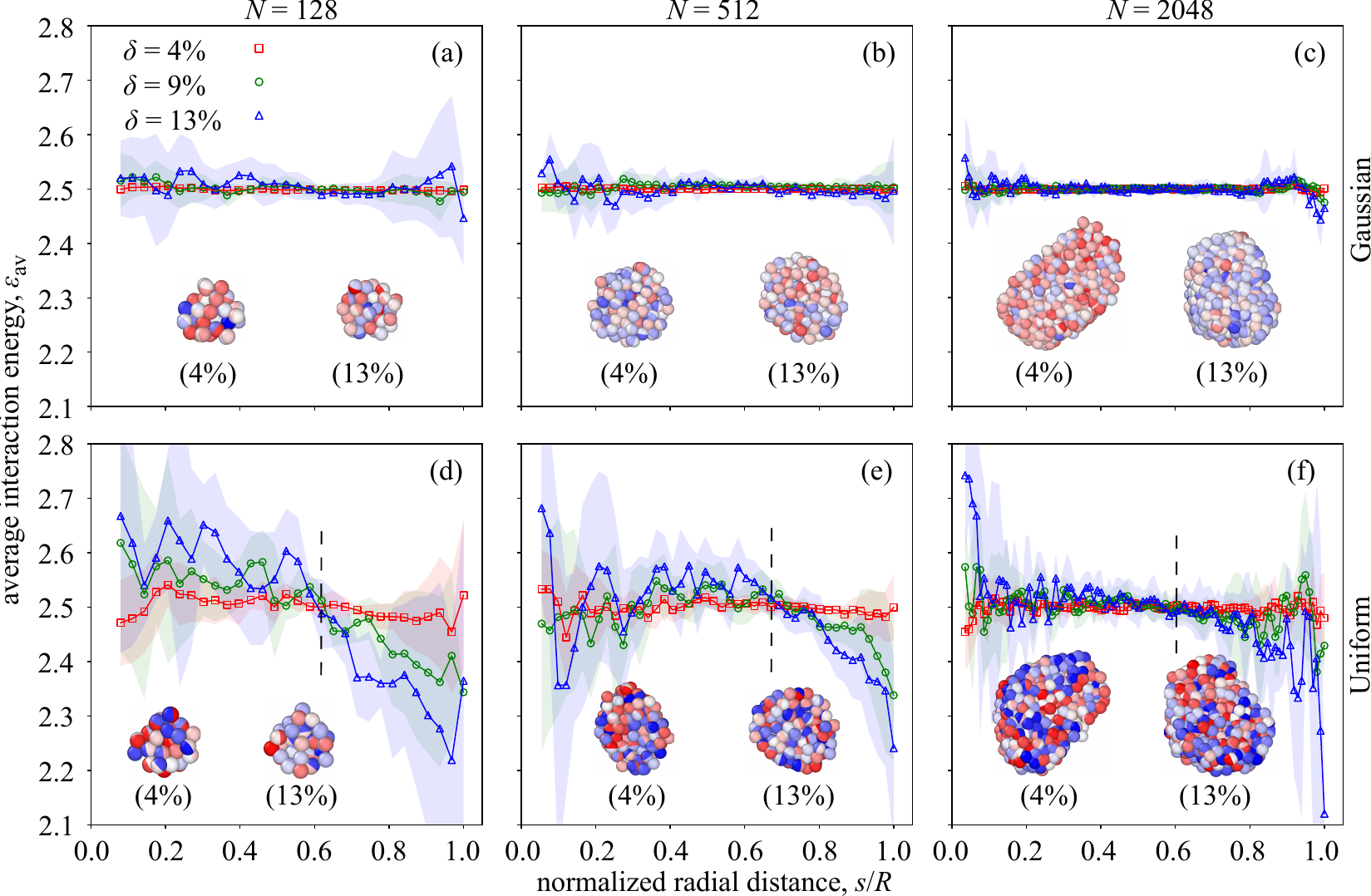}
		\caption{Plot of average interaction parameter, $\varepsilon_{\rm av}$, vs normalized radial distance, $s/R$ (with $R$ the globule radius), at different values of chain length $N$ indicated in the figure shown for Gaussian, figure (a)-(c), and uniform, figure (d)-(f), energy polydisperse chains. The shaded regions represent the errors. The data points are obtained by considering 1000 independent configurations from a replica and then averaged over 10 replicas for each $\delta$. Inset: Snapshot of globule (sliced view) to show the monomer organization and corresponding $\delta$ value (in \%) is also indicated.}
		\label{epsilion-s}
	\end{center}
\end{figure*}

\begin{figure*}[]
	\begin{center}
		\includegraphics*[width=0.95\textwidth]{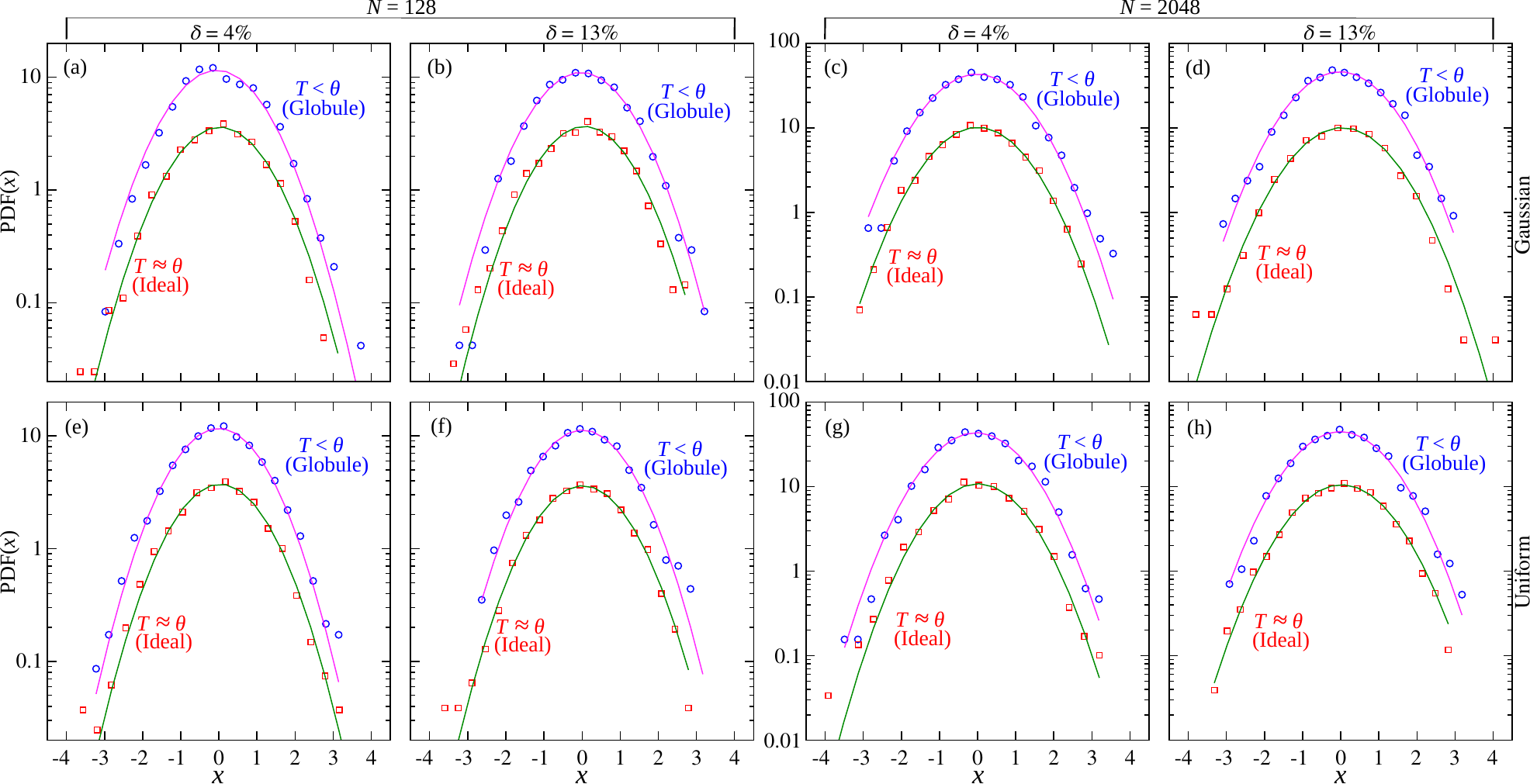}	
		\caption{The probability density function of rescaled potential energy per particle, $x=\left(u-\langle u \rangle \right)/\sigma_u$ where $\langle u \rangle$ is the average and $\sigma_u^2$ the variance, at $T \approx \theta^*$ and $T < \theta^*$ for both Gaussian (a-d) and uniform (e-h) energy polydisperse chains (of length $N=128~{\rm and}~2048$) show for $\delta=$ 4\% and 13\% indicated in the figure. The solid lines represent Gaussian fit to the respective data points.}
		\label{PE}	
	\end{center}
\end{figure*}

Since the monomers have distinct interaction parameters, $\varepsilon_{i}$ (drawn randomly from the given distribution), it is expected that, in the globular state, favorable contacts which lowers the energy of the systems form neighboring pairs. The favorable contacts are studied by means of contact map which, for any $i$-$j$ monomer pair, is defined as 
\begin{equation}
C_{ij} =
\begin{cases}
1, \quad {\rm if}~ r_{ij} \le r_{c1} \\
0, \quad {\rm otherwise}~,
\end{cases}
\end{equation} 
where $r_{ij}$ is the separation between the $i$-$j$ pair, and $r_{c1}=1.2\sigma$ is the cut-off distance for defining neighbors. Here, the value of $r_{c1}$ roughly corresponds to the first minima of pair correlation function. For a polymer chain of $N$ monomers, $C_{ij}$ represents the elements of a $N \times N$ connectivity matrix and corresponds to the energy of the nearest neighbour in contact.\cite{LMirny, Chan} \\

In figure~\ref{contact-map}, we display the 2D contact map (for $N=128,512$) obtained from 5000 independent configurations (of a particluar realization of random $\varepsilon_i$ values, i.e., a randomly choosen replica) at three different values of polydispersity index $\delta$ for both Gaussian and uniform systems. Apart from neighboring monomers along the chain (where the contact probability is 1) there are preferred monomer-monomer contacts as indicated by the non-zero value of probability at off-diagonal elements of the matrix. Static pattern of contact map is expected for polymer chain with specific interaction sites (or fixed primary sequence). However, in our case, the observed pattern differ for different realizations of random $\varepsilon_i$ values. It also depends on the polydispersity index. In order to understand an effective picture of monomer organization in the compact globule, we further calculate the mean contact probability defined as  
\begin{equation}
P_c(d) = \frac{1}{N - d} \sum_{i=0}^{N - d} P_{i,i+d}~,
\label{eq: mean-contact-prob}
\end{equation}
where $d = |i - j|$ is the monomer distance along the chain. In general, $P_c(d) \sim d^{-\phi}$ and the value of exponent $\phi$ characterizes the packing of chain in the globular state.\cite{Nazarov, Imakaev, Mirny, Dekker, Aiden} For example, based on simple mean-field arguments, in a fractal space-filling like structure $\phi=1$ and thus $P_c(d) \sim d^{-1}$.\cite{Aiden} However, a more accurate study (which goes beyond the mean-field approximation) reveal that $\phi\approx 1.05 - 1.09$.\cite{Grosberg2014} In our study, for the energy polydisperse polymers, we calculate the mean contact probability and check the nature of packing.

In figure~\ref{fig: mean-contact-prob}, we display the mean contact probability (obtained from 1000 independent configurations and averaged over 10 replicas) for the longest chain length considered in this study, i.e., $N=2048$. As one can see, for both Gaussian and uniform energy polydisperse chains, the scaling $P_c(d) \sim d^{-\phi}$ is observed with two distinct regimes: (i) $\phi \approx 2$ (or in the range $1.89 - 2.2$) for $d \le 10$ which is larger than that of a 3D random walk where $\phi=1.5$,\cite{Imakaev} and (ii) for $10 \le d \le 100$ the exponent $\phi$ is closer to $1$ (or in the range $0.72 - 0.83$). (For $d>100$ the number of data points are relatively less and thus statistically not significant and therefore the discussion is omitted here.)  
%Interestingly, the range in which $P_c(d) \sim d^{-1}$ increases with $N$. \textcolor{blue}{[MORE DISCUSSION]} \\

Another interesting aspect of packing is the spatial organization of energetically different monomers within the globule which we characterize by the radially averaged interaction parameter, $\varepsilon_{\rm av}= \sum_{i=1}^{n_s} \varepsilon_{\rm i}/n_s$ with $n_s$ as the number of particles in a shell of thickness d$s$ at a radial distance $s$ from the center of the globule. In figure~\ref{epsilion-s}, variation of $\varepsilon_{\text{av}}$ with normalized radial distance is shown for both Gaussian and uniform systems for different chain lengths and polydispersity index $\delta$. 
In case of the Gaussian system, the average value $\varepsilon_{\text{av}} \approx \langle\varepsilon\rangle$ throughtout the globule indicating a homogeneous (in terms of particle identity) distribution within the globule; however, a slight increase (or decrease) at the inner core (or edge) of the globle is visible for large chain length at $\delta=13\%$ (not visible for smaller value of $\delta$). 
On the other hand, for the uniform system, it is observed that $\varepsilon_{\text{av}} > \langle\varepsilon\rangle$ for distance $s \le 10\%R$ (inner core) of the globule size, and for $10\%R < s <60\%R$ (mid-region) the value of $\varepsilon_{\text{av}} \approx \langle\varepsilon\rangle$ (oscillating about the mean value) and for larger distance, i.e. $s \ge 60\%R$ (outer-region), the value is lower than the mean (which is independent of chain length).  
The overall trend indicates monomer segregation based on the identity, i.e., the inner core of the globule predominently consists of particles with high $\varepsilon_i$ values, whereas in the mid-region there is homogeneous particle distribution, and the outer-region consists mostly of low $\varepsilon_i$ particles. The observed self organization or enrichment (or depletion) of high $\varepsilon_i$ particles at the core (or edge) of the globlue is more pronounced for larger value of $\delta$. Similar particle identity ordering is also observed in energy polydisperse fluid systems, where as a consequence of segregation interesting particle dynamics emerges.\cite{shagolsem2015,shagolsem2016} The difference between the Gaussian and the uniform systems can be understood from the fact that, for a given value of $\delta$, the width of $\varepsilon_{ij}$ (or $\varepsilon_{i}$) distribution is relatively large for the uniform system, see figure~\ref{fig: eps-ij-distribution}, and consequently population of the particles that belong to the tail of the distribution is large leading to the pronounced effect. \\

In the fluctuations of a wide variety of systems, e.g. in turbulence and magnetic systems, it is observed that as one approaches the critical point, the distribution of the fluctuations deviates from the Gaussian behavior.\cite{bramwell1998,portelli2001} Here also we check whether any such deviation is seen by calculating the distribution of the fluctuations of global potential energy per particle $u=U/N$ (with $U$ as the total potential energy, see equation~\ref{eqn: LJ-potential}, and $N$ as the number of monomers) when the chain adopts ideal statistics and in the globule state, i.e., at $T\approx \theta^*$ and $T < \theta^*$, respectively. The rescaled distributions (with zero mean and unit variance) of potential energy is shown in figure~\ref{PE} for both Gaussian and uniform systems for two values of polydispersity index $\delta$. At both the considered points, the fluctuation is well described by the Gaussian curve and no significant deviation is observed. In the following section, we investigate the time evolution of polymer conformation leading to globular state by means of shape measures and mean contact probability.   

%%%%%%%%%%%%%%%%%%%%%%%%%%%%%%%%%%%%%%%%%%%%%
%%%%%%%%%%%%%%%%%%%%%%%%%%%%%%%%%%%%%%%%%%%%%

\subsection{Folding Kinetics} 
\label{sec: folding-kinetics}

\begin{figure*}[ht]
	\begin{center}
		\includegraphics*[width=0.97\textwidth]{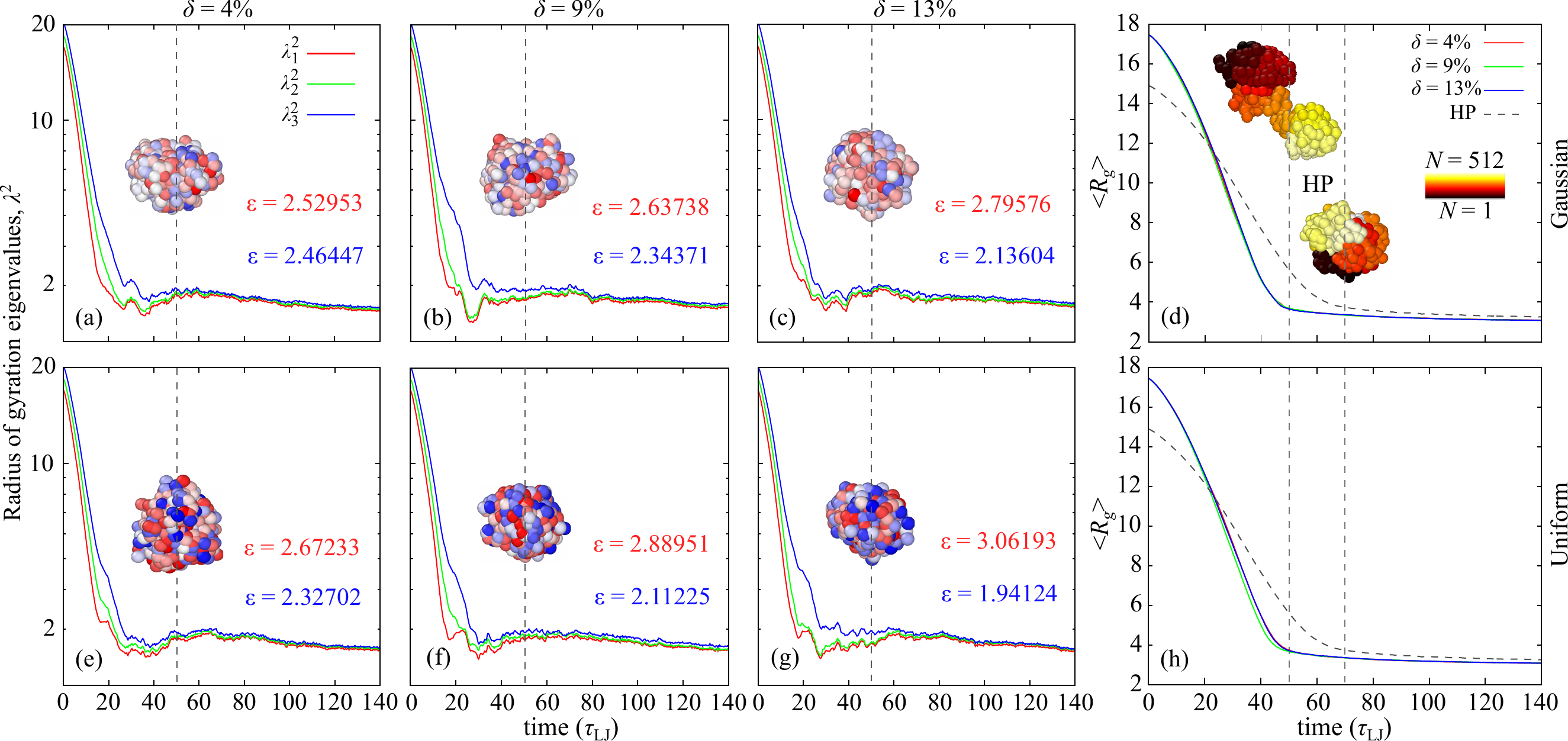}
		\caption{Time evolution of the eigenvalues, $\lambda_1^2$, $\lambda_2^2$ and $\lambda_3^2$ of the gyration tensor and $\langle R_g \rangle$ for Gaussian (G) and uniform (U) systems at $\delta$ = 4\%, 9\%, 13\%. Homopolymers (HP) are represented by the dashed linesin in figure (d) and (h).}
		\label{V-time}
	\end{center}
\end{figure*}

Conformational changes during folding from an extended state to compact globular structure is monitored through gyration tensor {\bf S} as instantaneous shape measures.\cite{theodorou1985,solc_stockmayer_1971, solc1971} By choosing a principal axis system, where {\bf S} is in a diagonal form with eigenvalues $\lambda_1^2, \lambda_2^2$ and $\lambda_3^2$ such that $\lambda_1^2 \le \lambda_2^2 \le \lambda_3^2$, the instantaneous shape of a polymer chain is characterized in terms of asphericity, $b$, and acylindricity, $c$, defined as   
	\begin{align}
	b &= \lambda_3^2 - \frac{1}{2} (\lambda_1^2+\lambda_2^2)~,~{\rm and}\nonumber \\
	c &= \lambda_2^2 - \lambda_1^2~.
	\label{eqn:shape-param}
	\end{align}
\noindent Here, the quantity $b$ and $c$ measure the deviations from spherical and cylindrical shapes, respectively, with $b=0$ and $c=0$ for perfect sphere and cylinder, respectively. And trace of {\bf S} represent the squared radius of gyration, i.e., $\langle R_g^2 \rangle = s^2 = \lambda_1^2+\lambda_2^2+\lambda_3^2$ . \\
To this end, the system initially at high temperature ($T_0=20$) is quenched to $T_0=1$ and the evolution of eigenvalues and average size are monitorted, see figure~\ref{V-time} for a chain of $N=512$. It is clear from the figure that coil-globule transition happens at a relatively short time scale, e.g., by $t\approx 50\tau_{_{\rm LJ}}$, the eigenvalues are roughly constant and as time increases difference between the eigenvalues are quite small and the structure is already a globule. The same is true for a homopolymer; however, the globule is achieved at a slightly later time of $t \approx 70 \tau_{_{\rm LJ}}$, see figure~\ref{V-time}(d),(h) (also see supplementary information). Apparently, no significant effect on the transient phase is observed as a consequence of having different polydispersity index (within the considered range of values) for both Gaussian and uniform systems.\\

\begin{figure}[h]
	\begin{center}
		\includegraphics*[width=0.47\textwidth]{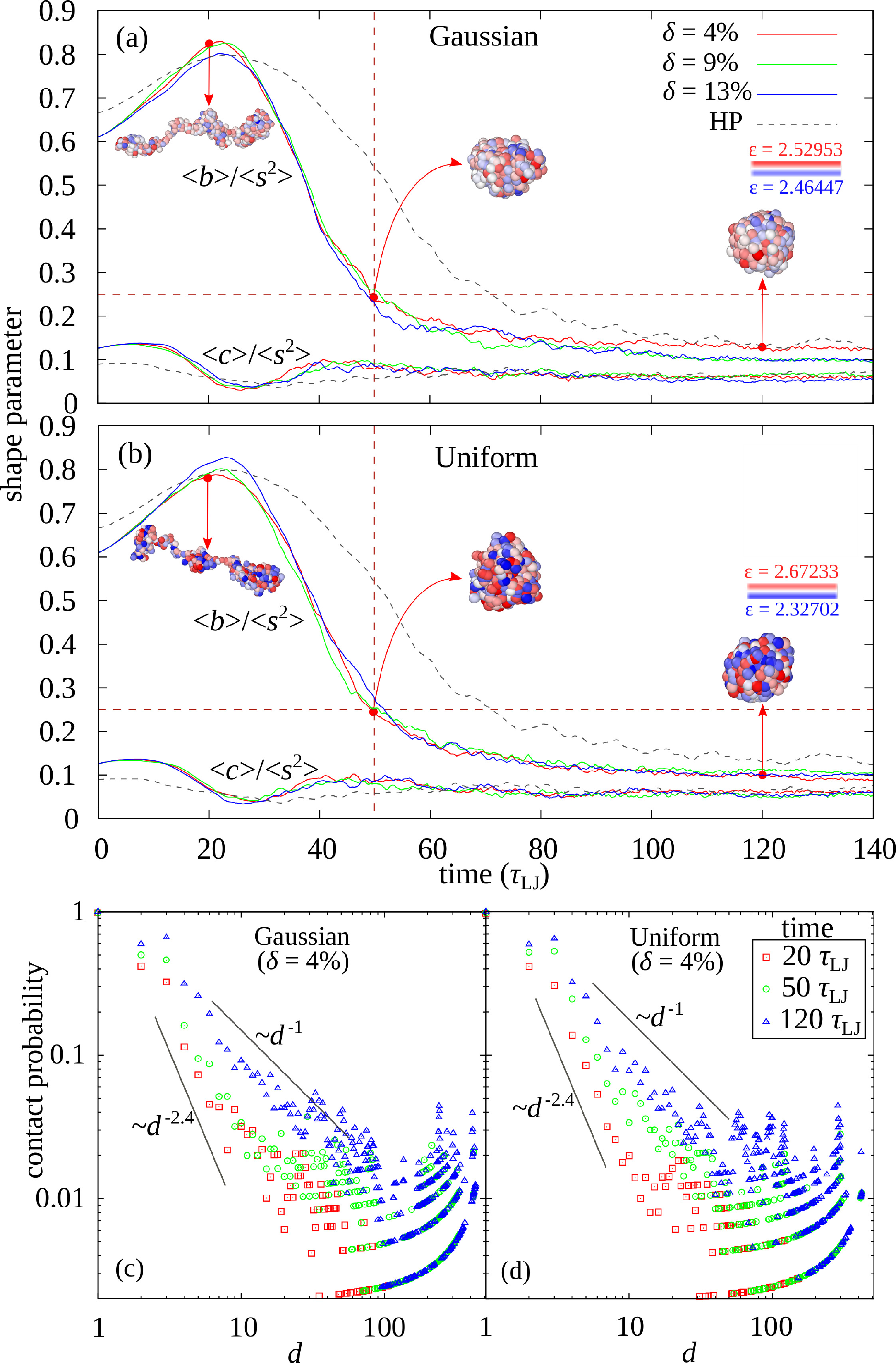}
		\caption{The evolution of normalised asphericity, $ \langle b \rangle$ and acylindricity, $ \langle c \rangle$ following quenching for chain of $\delta = 4\%, 9\%, 13\%$. Vertical and horizontal dashed lines are guide to the eye to compare the values at $t=50\tau_{_{\rm LJ}}$. (c)-(d): Instantaneous mean contact probabilities at different times are indicated in the figure (shown for $\delta$ = 4\%).}
		\label{b-c-time}	
	\end{center}
\end{figure}

The evolution of average value of asphericity, $\langle b \rangle$, and acylindricity, $\langle b \rangle$, normalized by $\langle s^2 \rangle$ is shown in figure~\ref{b-c-time}(a)-(b) for both energy polydisperse systems plotted along with that of the homopolymer. As we can see in the figure, for both Gaussian and uniform systems, immediately following the quench, $\langle b \rangle / \langle s^2 \rangle$ increases from an initial value of about 0.6 to a maximum of about 0.8 when $t\approx 20\tau_{_{\rm LJ}}$, where polymer adopts pearl-necklace conformation (an intermediate state during folding), which is followed by a rapid decrease and then a saturation around 0.1 at sufficiently long time. On the other hand, the value of $\langle c \rangle / \langle s^2 \rangle$, after remaining roughly constant (around 0.15) for a while, decreases to a minimum of about 0.05 when $t\approx 25\tau_{_{\rm LJ}}$ which is followed by a slight increase and then the saturation occurs. On comparison, the overall change in the value of acylindricity is small. It is interesting to note that, compared to homopolymer, energy polydisperse chain reach globular state rather rapidly. For instance, at $t\approx 50\tau_{_{\rm LJ}}$, for energy polydisperse chains $\langle b \rangle / \langle s^2 \rangle \approx 0.25$, while for homopolymer it is roughly 0.53, see figure~\ref{b-c-time}, also see figure~\ref{V-time}(d) for the relative difference in the chain conformations. \\

Finally, we check the evolution of instantaneous mean contact probability at 20 (early), 50 (intermediate), and 120 (late) time in $\tau_{\text{LJ}}$ units during the folding. The contact probabilities obtained from different replicas are shown in figure~\ref{b-c-time}(c)-(d) for both Gaussian and uniform energy polydisperse chains at $\delta \approx 4\%$. We observe that $P_c(d) \sim d^{\phi}$ for both systems, and during the transient phase the magnitude of exponent decreases from $\phi \approx -2.4$ (in the extended state) to $\phi \approx -1$ (in the stable globular state) indicating formation of fractal-like globular structure as shown in the figure.

%%%%%%%%%%%%%%%%%%%%%%%%%%%%%%%%%%%%%%%%%%%%%%%%%% 
%%%%%%%%%%%%%%%%%%%%%%%%%%%%%%%%%%%%%%%%%%%%%%%%%%

\section{Conclusion}
\label{sec: summary}

Coil-globule transition is a classic problem which is still of great interest for its relevance to protein folding related problems. In this study also, we have carried out an extensive computer simulation of an energy polydisperse chain, a model heteropolymer (belonging to the independent interaction model), and investigated the effect of the functional form of the distribution and polydispersity index, $\delta$, on the coil-globule transition in general. In particular, considering a Gaussian and uniform distribution functions, the consequences of varying $\delta$ (up to 13\%) on the $\theta$-temperature, chain size scaling, identity organization in the globular state, and kinetics of the folding are addressed in this study. \\ 
It is observed that, for a given distribution, with increasing $\delta$ (or variance of the distribution) the $\theta$-temperature increases in general and its $N$ dependence follows $\theta(N)\approx\theta^\ast(1-1/\sqrt{N})$. As one would expect, the chain size $R_g\sim N^\nu$ with $\nu\approx 0.5$ at the respective $\theta$-temperature for all values of $\delta$ considered. Interestingly, the asymptotic value $\theta^\ast$ for the model heteropolymer is found to be lower than the homopolymer counterpart and they follow the sequence: $\theta^\ast({\rm U}) < \theta^\ast ({\rm G}) < \theta^\ast ({\rm HP})$. For the independent interaction model,\cite{Pande1997} the transition temperature is shown to be proportional to the variance of the distribution which is in agreemenent with our result. Furthermore, our finding that $\theta^\ast({\rm U}) < \theta^\ast ({\rm G})$ suggest that the value of $\theta^\ast$ is dictated by the most probable value, $\varepsilon_{ij}^{\rm mp}$, of the distribution of pair-wise interaction rather than the width of the distribution, i.e., at a given $\delta$, here $\varepsilon_{ij}^{\rm mp}({\rm U}) < \varepsilon_{ij}^{\rm mp}({\rm G})$ despite uniform distribution being broader (see figure~\ref{fig: eps-ij-distribution}). However, the nature of collapse transition turns out to be universal in the sense that when scaled properly all the swelling curves fall on a master curve, i.e., it is independent of the functional form and variance of the distribution, which is also not different from that of the homopolymer. \\

The 2D contact map, despite showing favourable monomer-monomer contacts in the compact state, lacks a specific pattern due to the random nature of interaction sites along the chain. However, the internal polymer organization characterized by the mean contact probability show distinct regimes: (i)  at short countour distance, $d \le 10$, we find $P_c(d) \sim d^{-\phi}$ with $\phi: 1.89 -2.2$, and (ii) for $d$ in the range $10-100$ we get $\phi: 0.72 - 0.83$. On the other hand, spatial organization of monomers in the compact state show segregation based on the monomer identity. It is observed that monomers with high interaction parameters are predominently at the core (typically less than 10\% of the globule radius, where the average interaction parameter is larger than the mean of the distribution) followed by a region of homogeneous distribution and eventually decays below the mean for distance beyond roughly 60\% of the globule radius. Relatively, the segregation is more pronounced in the case of uniform distribution and it increases with higher $\delta$. Finally, the study of structural evolution following quenching revel that coil to globule transition is faster in the case of energy polydisperse chain (and weakly depend on $\delta$). \\
Since topology and shapes of proteins play an important role in many biological functions understanding the interplay of topology and instantaneous shape of this model heteropolymer is important. Therefore, as an extension to the current study we shall explore the effect of having different topologies on the phase transition and other static properties of the model heteropolymer system. 
%And in addition, effect of considering explicit solvent, behaviour in a confinement, rheology and dynamics are yet to be explored. 
%\clearpage
%%%%%%%%%%%%%%%%%%%%%%%%%%%%%%%%%%%%%%%%%%%%%%%%%%%%%%%%
%%%%%%%%%%%%%%%%%%%%%%%%%%%%%%%%%%%%%%%%%%%%%%%%%%%%%%%%

\begin{acknowledgements}
T.~Vilip gratefully acknowledges fruitful discussions with M.~Premjit, J.~Pame, T.~Premkumar, S.~Jimpaul, and U.~Somas. %L.S.S. acknowledge the financial support through DST-INSPIRE Faculty Award.
\end{acknowledgements}

%\clearpage

%\begin{figure}[h]
%	\begin{center}
%		\includegraphics*[width=0.4\textwidth]{figures/fig-EPP-ToC.png}
%		\caption{Graphical abstract figure.}
%		%\label{zzz}	
%	\end{center}
%\end{figure}

\end{document}